\pdfoutput=1

\documentclass[11pt]{article}

\usepackage[]{coling}

\usepackage{times}
\usepackage{latexsym}
\usepackage{rotating}

\usepackage[T1]{fontenc}

\usepackage[utf8]{inputenc}

\usepackage{microtype}

\usepackage{inconsolata}

\usepackage{graphicx}

\usepackage{color}

\usepackage{booktabs}
\usepackage{multirow}
\usepackage{graphicx}
\usepackage{colortbl}
\usepackage{tikz}
\usepackage{url}
\usepackage{forest}
\usetikzlibrary{trees,positioning,shapes,shadows,arrows.meta}

\usepackage{booktabs}
\usepackage{tabularx} 
\usepackage{lscape} 
\usepackage{amssymb}
\usepackage{pifont}

\definecolor{indigo}{HTML}{4B0082}
\definecolor{amber}{HTML}{cc4e00}
\definecolor{teal}{HTML}{008080}
\definecolor{grey}{HTML}{979ea8}

\title{From Generalist to Specialist: A Survey of \\  Large Language Models for Chemistry}

\author{Yang Han$^{1,2}$ ~~Ziping Wan$^{2}$ ~~Lu Chen$^{1,2*}$  ~~Kai Yu$^{1,2}$ ~~Xin Chen$^{2}$\thanks{\ \ Lu Chen and Xin Chen are the corresponding authors.} \\
$^{1}$X-LANCE Lab, Department of Computer Science and Engineering \\
MoE Key Lab of Artificial Intelligence, SJTU AI Institute \\Shanghai Jiao Tong University, Shanghai, China \\
$^{2}$Suzhou Laboratory, Suzhou, China \\
{\tt \{csyanghan,chenlusz\}@sjtu.edu.cn,mail.xinchen@gmail.com}
}

\begin{document}
\maketitle
\begin{abstract}
Large Language Models (LLMs) have significantly transformed our daily life and established a new paradigm in natural language processing (NLP). 
However, the predominant pretraining of LLMs on extensive web-based texts remains insufficient for advanced scientific discovery, particularly in chemistry. 
The scarcity of specialized chemistry data, coupled with the complexity of multi-modal data such as 2D graph, 3D structure and spectrum, present distinct challenges. 
Although several studies have reviewed Pretrained Language Models (PLMs) in chemistry, there is a conspicuous absence of a systematic survey specifically focused on chemistry-oriented LLMs.
In this paper, we outline methodologies for incorporating domain-specific chemistry knowledge and multi-modal information into LLMs, we also conceptualize chemistry LLMs as agents using chemistry tools and investigate their potential to accelerate scientific research.
Additionally, we conclude the existing benchmarks to evaluate chemistry ability of LLMs. Finally, we critically examine the current challenges and identify promising directions for future research.
Through this comprehensive survey, we aim to assist researchers in staying at the forefront of developments in chemistry LLMs and to inspire innovative applications in the field.
\footnote{We maintain an up-to-date Github repository at: \url{https://github.com/OpenDFM/LLM4Chemistry}.} 
\end{abstract}

\section{Introduction}
\label{intro}

\begin{figure}
    \centering
    \includegraphics[width=1\linewidth]{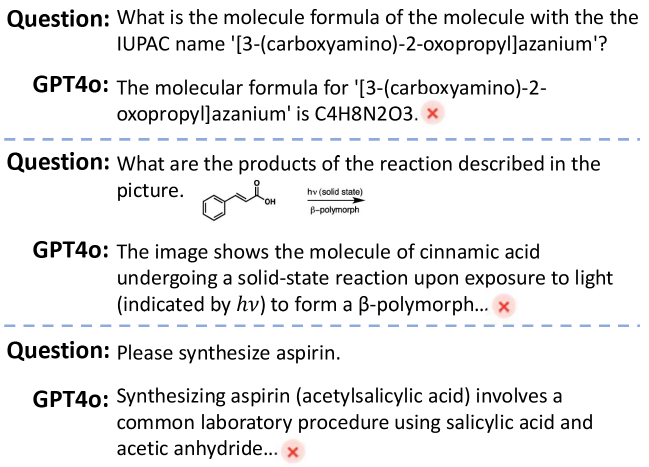}
    \caption{Three common errors in general LLMs arising from the key challenges.}
    \label{fig:examples}
\end{figure}

Recent years have witnessed remarkable advancements in daily life driven by LLMs. Competitive models like GPT-4 \cite{achiam2023gpt} and Claude \cite{claude2024} have demonstrated exceptional abilities across diverse tasks, often matching or surpassing human-level performance, marking significant progress toward Artificial General Intelligence (AGI,~\citet{bubeck2023sparks}).
In scientific domains, LLMs have been applied to handle tasks involving natural language and various scientific data (\textit{e.g.}, molecules, proteins, DNA), showing promising results \cite{fang2023mol}. Among these, chemistry LLMs, further tailored for chemical applications via additional training or advanced prompt engineering, have garnered significant attention. 
Before the advent of LLMs, there are lots of notable efforts towards chemistry, such as MolT5 \cite{edwards2022translation}, Text2Mol \cite{edwards2021text2mol}, MoMu \cite{su2022molecular}, Text+Chem T5 \cite{christofidellis2023unifying}.
However, these models are built on PLMs like BERT \cite{devlin2018bert} and T5 \cite{raffel2020exploring}, requiring fine-tuning for specific tasks and lacking emergent abilities \cite{wei2022emergent}, such as Chain-of-Thought (CoT, \citet{wei2022chain}) reasoning and tool-using capabilities \cite{qin2023toolllm}.
Existing reviews \cite{xiao2024bridging, liao2024words, pei2024leveraging} have already discussed those PLMs in chemistry, such as \citet{liao2024words}, which emphasize molecule encoding methods and pretraining objectives. More related works are discussed in the Appendix \ref{related_work}. In contrast, \textbf{our survey focuses on generative models with Transformer decoder architectures \cite{vaswani2017attention}}, addressing key challenges of general LLMs and reviewing existing approaches to adapt them for chemistry-specific tasks and applications.

\begin{figure*}
    \centering

\tikzset{
    basic/.style  = {draw, text width=2cm, align=center, rectangle, font=\scriptsize},
    root/.style   = {basic, draw=grey, rounded corners=2pt, thin, align=center, fill=none},
    infnode/.style = {basic, draw=indigo, thin, rounded corners=2pt, align=left, fill=none, text width=2.2cm, align=center},
    tranode/.style = {basic,draw=amber, thin, rounded corners=2pt, align=center, fill=none,text width=2.2cm},
    valnode/.style = {basic, draw=teal, thin, rounded corners=2pt, align=center, fill=none, text width=2.2cm},
    valpretracitenode/.style = {basic, draw=teal, thin, rounded corners=2pt, align=center, fill=none, text width=3.5cm},
    infcitenode/.style = {infnode, thin, align=left, fill=none, text width=65mm, font=\tiny},
    tracitenode/.style = {tranode, thin, align=left, fill=none, text width=65mm, font=\tiny},
    pretracitenode/.style = {tranode, thin, align=left, fill=none, text width=36mm, font=\tiny},
    valcitenode/.style = {valnode, thin, align=left, fill=none, text width=65mm, font=\tiny},
    edge from parent/.style={draw=black, edge from parent fork right}

}
\begin{forest} for tree={
    grow=east,
    growth parent anchor=west,
    parent anchor=east,
    child anchor=west,
    anchor=center,
    edge path={\noexpand\path[\forestoption{edge},->, >={latex}] 
         (!u.parent anchor) -- +(10pt,0pt) |-  (.child anchor) 
         \forestoption{edge label};}
}
[Chemistry LLMs, root,  l sep=6mm,
    [Challenge 3: Chemistry Tools (\S~\ref{sec:challenge3}), infnode,  l sep=6mm,
        [Embodied Robots (\S~\ref{sec:physical_environment}), infnode,l sep=5mm,
[{Coscientist~\cite{boiko2023autonomous},CLAIRify~\cite{yoshikawa2023large},ORGANA~\cite{darvish2024organa},}, infcitenode]
        ]
        [ML Models (\S~\ref{sec:ml_models}), infnode,l sep=5mm,
            [{ChatChemTS~\cite{ishida2024large}, ChemCrow~\cite{m2024augmenting}, ChatMOF~\cite{kang2024chatmof}, ChemReasoner~\cite{sprueill2024chemreasoner}}, infcitenode]]
        [Structured Knowledge Retrieval (\S~\ref{sec:structure_knowledge}), infnode,l sep=5mm,            [{ChemCrow~\cite{m2024augmenting},LLaMP~\cite{chiang2024llamp}, ChatGPT Chemistry Assistant~\cite{zheng2023chatgpt}, DRAK-K~\cite{liu2024drak},}, infcitenode]]
    ]
    [Challenge 2: Multi-modal Data: (\S~\ref{sec:challenge2}), tranode,  l sep=6mm,
        [Other Modalites (\S~\ref{sec:other_modal}), tranode,l sep=5mm,
            [{
            ChemVLM \cite{li2024seeing}, ChemDFM-X \cite{zhao2024chemdfmx} }, tracitenode
            ]
        ]
        [3D Structure (\S~\ref{sec:3D-structure}), tranode,l sep=5mm,
            [{ 
            3D-MoLM\cite{li2024towards} }, tracitenode
            ]
        ]
        [2D Graph (\S~\ref{sec:2d_graph}), tranode,l sep=5mm,
            [{InstructMol~\cite{cao2023instructmol},  HIGHT~\cite{chen2024hight}, MolTC~\cite{fang2024moltc}, MolCA~\cite{liu2023molca}, ReactXT~\cite{liu2024reactxt}, ICMA~\cite{li2024large}, MoleculeGPT~\cite{zhang2023moleculegpt}, MolX~\cite{le2024molx}, MM-RCR ~\cite{zhang2024text} }, tracitenode
            ]
        ]
        [1D Sequence (\S~\ref{sec:1d_sequence}), tranode,l sep=5mm,
            [{MolX~\cite{le2024molx},  MoleculeGPT~\cite{zhang2023moleculegpt}}, tracitenode
            ]
        ]
    ]
    [Challenge 1: Domain Knowledge (\S~\ref{sec:challenge1}), valnode,  l sep=6mm,
        [RLHF (\S~\ref{rlhf}), valnode, l sep=5mm,
            [
            {MolGen~\cite{fang2024domain}, BindGPT~\cite{zholus2024bindgpt}, MolRL-MGPT~\cite{hu2024novo}}, valcitenode
            ]
        ]
        [SFT (\S~\ref{sft}), valnode, l sep=5mm,
            [   Task-specific SFT, valnode, l sep=5mm,
                [
                {GPTChem \cite{jablonka2024leveraging}, MatChat~\cite{chen2023matchat}, MolecularGPT~\cite{liu2024moleculargpt}}, valpretracitenode
                ]
            ]
            [   Multi-task SFT, valnode, l sep=5mm,
                [
                {LlaSMol~\cite{yu2024llasmol}, Mol-Instructions~\cite{fang2023mol}, ChemDFM~\cite{zhao2024chemdfm}, ChemLLM~\cite{zhang2024chemllm}}, valpretracitenode
                ]
            ]
        ]
        [Pre-training (\S~\ref{pre-training}), valnode, l sep=5mm,
            [
                {ChemDFM~\cite{zhao2024chemdfm}}, valcitenode
            ]
        ]
    ]
]
\end{forest}
    \caption{Taxonomy of currect approachs for transfering general LLMs to specialized chemistry LLMs.}
    \label{fig_tax}
\end{figure*}
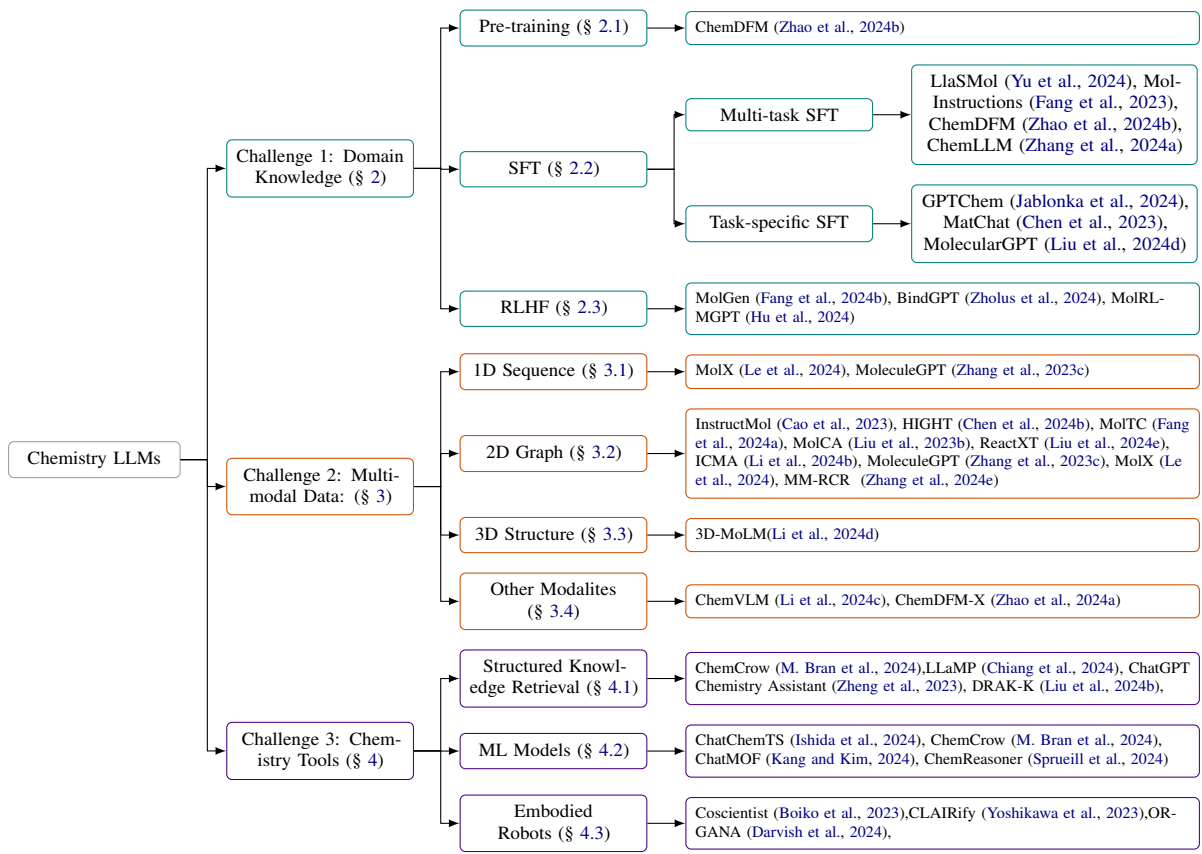

General LLMs, such as the GPT \cite{ouyang2022training, achiam2023gpt} and LLaMA series \cite{touvron2023llama1, touvron2023llama2}, have demonstrated impressive performance. However, they tend to underperform on chemistry-related tasks as shown in Figure \ref{fig:examples}. We identify three key challenges contributing to these limitations.

\textit{\textbf{Challenge 1: domain knowledge is not enough.}} Most LLMs are pre-trained with the objective of predicting the next token based on web data sourced from the internet \cite{ouyang2022training}, as demonstrated by open-source models like LLaMa series \cite{touvron2023llama1, touvron2023llama2}. While some chemistry-related data exist within these datasets, the quantity is minimal, and there is a lack of data specifically tailored for chemistry. This deficiency extends to other crucial steps in the development of LLMs, such as supervised fine-tuning (SFT) and reinforcement learning from human feedback (RLHF, \citet{christiano2017deep, stiennon2020learning}).

\textit{\textbf{Challenge 2: multi-modal data is not perceived.}} Chemistry encompasses various modalities, including 1D sequences \cite{krenn2020self}, 2D molecular graphs \cite{duvenaud2015convolutional, xu2018powerful, liu2019n}, and 3D structures \cite{schutt2018schnet, satorras2021n, atz2021geometric}. Additionally, there are numerous chemical spectra, such as Nuclear Magnetic Resonance (NMR, \citet{simpson2012nuclear}), Liquid Chromatography-Tandem Mass Spectrometry (LC-MS, \citet{seger2012usage, duhrkop2015searching, litsa2023end}), and Infrared Spectroscopy (IR, \citet{alberts2023leveraging}). These spectra contain substantial information that LLMs currently fail to fully exploit.

\textit{\textbf{Challenge 3: chemistry tools are not utilized.}} Due to the core design of LLMs, they often struggle with retaining up-to-date knowledge and performing specific chemistry operations \cite{castro2023large, schick2024toolformer}. On the other hand, there are numerous powerful chemistry tools, such as the structure knowledge retrieval (PubChem \cite{kim2019pubchem}, OPTIMADE \cite{evans2024developments}), and various expert-designed artificial intelligence systems tailored to address specific problems like reaction prediction \cite{pesciullesi2020transfer}, retrosynthesis planning \cite{segler2018planning} and so on. The absence of integration with these chemistry tools significantly hinders the performance of LLMs in the field of chemistry.

In this survey, we critically review current efforts addressing the three key challenges outlined in Figure \ref{fig_tax}. Additionally, we review the existing benchmarks used to evaluate the performance of chemistry LLMs and offer suggestions for future research directions. 
To the best of our knowledge, this is the first systematic survey reviewing existing approaches for transferring general LLMs to chemistry-specific LLMs in decoder architecture.
\section{Domain Knowledge}
\label{sec:challenge1}
Pre-training, SFT and RLHF have been the de facto way to enhance domain knowledge of LLMs. We will detail those methods in the following sections.

\subsection{Pre-training}
\label{pre-training}

The natural of LLMs lay in language modeling, given a set of examples $(x_1, x_2, ..., x_n)$ each composed of variable length sequences of symbols $(s_1, s_2, .., s_m)$, language model is framed as a unsupervised distribution estimation and the joint probabilities over symbols can be formulated \cite{radford2019language}:
\begin{equation}
    p(x) = \prod_{i=1}^{n} p(s_n | s_1, ..., s_{n-1}),
\end{equation}
self-attention architectures like the Transformer can be applied to compute these conditional probabilities.
Training on a large-scale corpus in this manner enables LLMs to capture rich language representations, refering to pre-training.

Continue pre-training is prefered given the existence of advanced foundation models like LLaMA~\cite{touvron2023llama1, touvron2023llama2} and Galactica \cite{taylor2022galactica}, which already contain some basic chemistry knowledge. In contrast, pre-training from scratch is cost-prohibitive.
Chemistry knowledge is typically encoded in specific languages, such as the Simplified Molecular-Input Line-Entry System (SMILES) \cite{weininger1988smiles}, which represents 3D structures as flattened sequences while preserving most structural information. Other representations include molecular formulas, SELFIES \cite{krenn2020self}, International Union of Pure and Applied Chemistry (IUPAC) names, and the Chemical Identifier (InChI) \cite{heller2013inchi}.
To enhance foundation models with domain-specific chemistry knowledge, it is necessary to gather pre-training corpora in these chemical languages and apply continued pre-training.


The volume of pre-training data required for chemistry LLMs is immense, making it difficult to obtain and, in some cases, restricted by copyright.
To the best of our knowledge, ChemDFM \cite{zhao2024chemdfm} is the sole chemistry LLM specifically pre-trained on a chemical corpus.
ChemDFM's training data comprises 34 billion tokens from 3.9 million chemical papers collected online before January 2022 and 49 million tokens from 1.4 thousand chemistry books sourced from LibreTexts\footnote{https://libretexts.org/} and Gold Books\footnote{https://goldbook.iupac.org/}.
Through pre-training on this chemical text, ChemDFM can acquire a solid understanding of chemistry and emerge as the top open-source model \cite{feng2024sciknoweval}.
Another T5-based chemistry LM, Nach0 \cite{livne2024nach0}, collects 13 million abstracts from PubMed, 119K patent descriptions from the USPTO, and incorporates approximately 100 million documents from ZINC.

\subsection{SFT}
\label{sft}

Pre-training on large corpus with next token prediction does not align well with users’ objective, as users expect models to "follow their instructions helpfully and safely" \cite{zhang2023instruction}. 
SFT effectively aligns LLMs with user expectations by training them on datasets consisting of (INSTRUCTION, OUTPUT) pairs, where INSTRUCTION refers to specific chemistry tasks and OUTPUT represents the desired responses.
Given the variety of chemistry tasks in the SFT dataset, it can be further categorized as follows:

\begin{enumerate}
    \item \textit{\textbf{Multi-task SFT}}: We categorize commonly used chemistry tasks into four types: SMILES understanding, reaction understanding, notation alignment and chemistry-related QA, as detailed in Appendix \ref{appendix:sft-task}.
    The most significant distinction among different SFT models \cite{yu2024llasmol, fang2023mol, zhao2024chemdfm, zhang2024chemllm} lie in their data sources and the volume of data used, and the detailed data distribution is shown in Appendix \ref{appendix:sft-task}.
    The total dataset volume ranges from 1.5M to 3M, although \citet{zhang2024chemllm} does not provide exact figures, it is likely of a similar magnitude.
    The distribution of tasks within the SFT dataset determines the model's chemistry capabilities, as identified by \citep{feng2024sciknoweval}. \citet{zhao2024chemdfm, zhang2024chemllm} focus more on chemistry-related QA, gathering major data from sources such as chemistry exams and existing datasets, which enhances the model's ability to answer user questions more naturally.

    \item \textit{\textbf{Task-specific SFT}}: Task-specific finetuning of LLMs has demonstrated effective prediction performances, often surpassing traditional machine learning models, particularly in low-data scenarios\cite{jablonka2024leveraging}.
    \citet{jablonka2024leveraging} finetune GPT-3 for classification, regression, and inverse design tasks, achieving competitive results in three case studies (polymers, metal-organic frameworks, and photoswitches). More recently, \citet{liu2024moleculargpt} propose hybrid instruction tuning on more than 1000 property tasks with LLaMA2-7b-chat \cite{touvron2023llama2}, reporting up to a 16.6\% average improvement over leading LLM baselines across all classification tasks. Additionally, \citet{chen2023matchat} also fine-tune LLaMA2-7B-chat with 13,878 pieces of structured material knowledge data to predict inorganic material synthesis pathways.

\end{enumerate}


In addition to these chemistry tasks, chemical text mining is also a crucial foundation in chemical research, as much scientific knowledge is dispersed across the text, tables, and figures in millions of academic papers \cite{dagdelen2024structured}. \citet{dagdelen2024structured} focus on joint named entity recognition and relation extraction, enabling the generation of simple English sentences or more structured formats, such as JSON object, from individual sentences or entire paragraphs. \citet{zhang2024fine} extend these efforts to more chemical text mining tasks, achieving the best performance across all tasks, with exact accuracy ranging from 69\% to 95\% using minimal annotated data.

\subsection{RLHF}
\label{rlhf}

While pre-training and SFT provide chemistry LLMs with domain-specific knowledge and enable them to perform specific tasks, these models are still prone to hallucination.
RLHF is the most effective method to alleviate hallucinations and build a truthful, helpful and harmless LLM \cite{ouyang2022training}. There are many detail algorithms to utilize human feedback, such as PPO~\cite{schulman2017proximal}, DPO~\cite{rafailov2024direct}. Beyond human feedback, other methods for collecting preference feedback include AI feedback \cite{lee2023rlaif, bai2022constitutional} and environment feedback \cite{cao2024towards, dong2024self}.


Existing research on human alignment for chemistry LLMs primarily focuses on molecular generation tasks.
\citet{fang2024domain} first pre-trains LLM on SELFIES \cite{krenn2020self}, enabling the generation of syntactically correct molecules; however, the model also produces undesirable molecules, referred as molecular hallucinations. To mitigate these hallucinations and better align with actual chemical contexts, they apply a rank loss \cite{liu2022brio} by assigning higher probabilities to molecule candidates with desired properties.
\citet{zholus2024bindgpt} finetunes a GPT-based model for 3D molecular design, and utlizes external feedback from docking software using REINFORCE algorithm \cite{williams1992simple}.
\citet{hu2024novo} further investigates multiple GPT agents to generate desirable molecules in diverse directions, with the reward function estimated by docking software. The objective is to maximize the average reward while simultaneously improving molecular diversity.

AI and environment feedback are the most commonly used rewards for chemistry LLMs, as the more valuable human feedback is often unavailable due to the need for strong domain knowledge and the lack of effective tools to collect chemistry-specific feedback. \citet{hu2024novo} design a Python-based open-source graphical user interface (GUI) to explore and evaluate molecules, and capture chemist’s implicit knowledge and preferences more efficiently. This tool provides a promising approach for collecting chemistry-specific feedback to better align chemistry LLMs with human expertise.

\section{Multi-Modal Data}
\label{sec:challenge2}

\begin{figure*}[]
\centering
\includegraphics[width=0.94\linewidth]{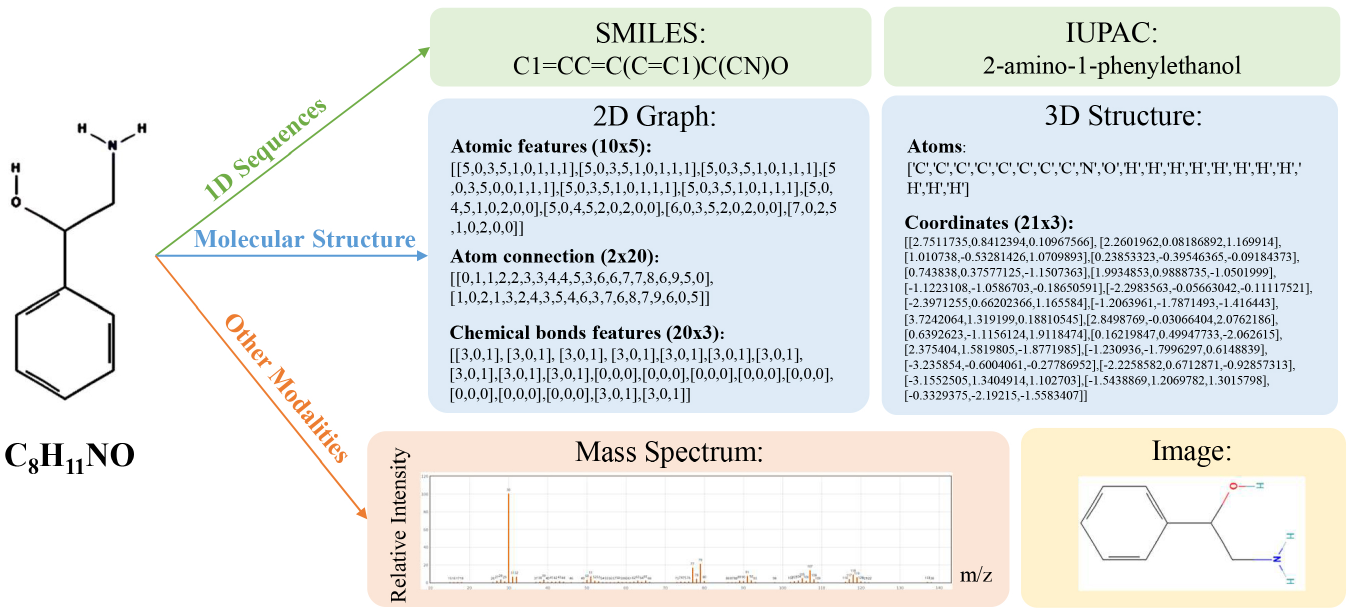} 
\caption{For example, the compound $C_{8}H_{11}NO$ can be represented across various modalities. 1D sequeues include SMILES, IUPAC name and so on. Molecular structure consist of 2D graphs and 3D structures, 2D graphs encompass three matrices: atomic features, atom connection, chemical bonds features, 3D strutures compromise the coordinate of every atom. Other modalities consist of mass spectra, images, and so on.}
\label{fig:modals}
\end{figure*}

Domain knowledge training is a standard approach for developing domain-specific LLMs, as demonstrated in fields like geoscience \cite{deng2024k2}, law \cite{zhou2024lawgpt}, and medicine \cite{zhang-etal-2023-huatuogpt}.
However, chemical data is highly fragmented across multiple modalities \cite{mirza2024bridging}, such as 2D graphs, 3D structures, and spectra, as shown in Figure \ref{fig:modals}, which cannot be directly processed by vanilla LLMs.
Inspired by recent advances in multi-modal and vision LLMs \cite{liu2024visual, li2024llava, huang2024language}, numerous studies have focused on integrating chemical modalities with vanilla LLMs through the design of alignment components. We provide a comprehensive review of these works based on the modalities they support: \textit{\textbf{1D Sequences}}, \textit{\textbf{2D Graphs}}, \textit{\textbf{3D Structures}}, and \textit{\textbf{Other Modalities}}.

\subsection{1D Sequences}
\label{sec:1d_sequence}

SMILES \cite{weininger1988smiles} is a widely used molecular representation, but it is generally processed as text using a byte-pair encoding tokenizer \cite{sennrich2015neural}, which fails to capture its inherent chemical information.
To address this limitation, MolX \cite{le2024molx} treats SMILES as a distinct modality and proposes a pre-trained BERT-like \cite{devlin2018bert} SMILES encoder to extract features, which are then aligned with other modalities through projection.
MoleculeGPT~\cite{zhang2023moleculegpt} also adapt ChemBerta \cite{ahmad2022chemberta} for SMILES encoding. 
However, SMILES lacks robustness and does not fully capture spatial information, leading to the development of other 1D sequence representations, such as SELFIES \cite{krenn2020self}, IUPAC names, molecular fingerprints \cite{morgan1965generation}, and InChI \cite{heller2013inchi}. These 1D sequences are generally processed similarly to text but can be further refined using specialized encoders, such as SELFormer \cite{yuksel2023selformer} for SELFIES and variational autoencoders (VAE, \citet{kingma2013auto}) for molecular fingerprints.

\subsection{2D Graphs}
\label{sec:2d_graph}

Compared to 1D sequences, 2D graphs offer a more intuitive representation of molecular structures and chemical bonds.
To process 2D graphs, an encoder is required to convert them into vector representations, followed by a projector to align these vectors with LLMs.
Graph neural networks (GNNs, \citet{hu2019strategies, xiao2022decoupled}) are widely used as 2D graph encoders and have been adopted by most multimodal chemistry LLMs \cite{liu2024reactxt, li2024large, zhang2023moleculegpt, le2024molx, zhang2024text}.
For instance, MolTC \cite{fang2024moltc} train two GNN-based encoders and representation projectors by freezing the LLM and backpropagating the generation loss.
InstructMol\cite{cao2023instructmol} employs MoleculeSTM’s graph encoder \cite{liu2023multi}, which is trained through molecular-textual contrastive learning.
MolCA \cite{liu2023molca} utilze a more expressive GNN model - Graph isomorphism network (GIN, \citet{hu2019strategies}), which pre-trained on 2 million molecules from the ZINC15 \cite{sterling2015zinc}.
HIGHT\cite{chen2024hight} further introduce a hierarchical graph tokenizer which em Vector Quantized-Variational AutoEncoder (VQVAE, \cite{zang2023hierarchical}) to extract high-order structural information and then feed them into LLMs.

There are various projectors to map graph features into the LLM embedding space, such as cross-attention \cite{alayrac2022flamingo}, Q-Former \cite{li2023blip}, position-aware vision language adapters \cite{bai2023qwen}, and light-weight Multi-layer Perceptron (MLP).
Q-Former is the most widely adopted projector \cite{liu2023molca, fang2024moltc, zhang2023moleculegpt}, maintaining a set of learnable query tokens to interact with the graph encoder and extract features. However, InstructMol~\cite{cao2023instructmol} argues that Q-Former requires a large number of paired data for pretraining, making alignment inefficient, and instead employs a lightweight MLP for alignment.
DeCo \cite{yao2024deco} also find that Q-Former tends to lose fine-grained visual attributes and spatial locality in visual LLMs.

\subsection{3D Structures}
\label{sec:3D-structure}

The 3D structures of molecules is crucial because it contains spatial information essential for understanding molecular dynamics, protein-ligand interactions, enzymatic functions, and other biomolecular phenomena \cite{li2024towards}. Unlike 1D sequences or 2D graphs, 3D structures provide a complete geometric representation of the molecule, allowing models to take into account the three-dimensional arrangement of atoms and the distances between them. 
MolLM \cite{tang2024mollm} and Uni-Mol \cite{zhou2023uni} demotarte performance enhancement in downstream tasks when incorporating 3D information.
3D-MoLM \cite{li2024towards} utilizes Uni-Mol \cite{zhou2023uni} to encode 3D conformations generated from SMILES and employs Q-Former \cite{li2023blip} for cross-modal alignment. This approach outperforms baseline models that rely on 1D or 2D molecular perceptions in tasks such as molecule-text retrieval, molecule captioning, and open-text question answering, particularly when addressing 3D-dependent properties.
In contrast, 3D-MolT5 \cite{pei20243d} contends that the modality alignment approach employed by 3D-MoLM \cite{li2024towards} is inefficient and introduces a specialized 3D vocabulary to train 1D, 3D, and text modalities within a unified architecture, demonstrating significant improvements over 3D-MoLM \cite{li2024towards} in various downstream tasks.

\subsection{Other Modalities}
\label{sec:other_modal}

2D graphs or 3D structures generated by RDKit are often represented as matrices, which are not human-readable. In contrast, chemical images are more intuitive and frequently used to represent chemical structures in a human-friendly format. At the same time, numerous efficient image algorithms, such as the Vision Transformer (ViT) \cite{dosovitskiy2020image} and Swin Transformer \cite{liu2021swin}, can be directly employed as modality encoders.
GIT-Mol \cite{liu2024git} utilizes Swin Transformer \cite{liu2021swin} from SwinOCSR for image ecoding, and adopt cross-attention for modal alignment.
ChemVLM \cite{li2024seeing} adopts InternViT-6B \cite{chen2024internvl} as the vision encoder, following the LLaVA \cite{liu2024visual} architecture in the "ViT-MLP-LLM" style.
Additionally, ChemVLM introduces three new chemical image datasets — ChemOCR, MMCR-Bench, and MMChemBench, However, these datasets are not open-source at this time. To facilitate future research on chemical images, we provide a summary of existing chemical image datasets in Appendix \ref{appendix:molecule_image_dataset}.



Another important chemistry-specific modality is spectral , which can be obtained through simulations (CFMID 4.0,~\citet{wang2021cfm}) and experiments. This data is rich in structural information and plays a vital role in determining molecular structures. For example, MSNovelist \cite{stravs2022msnovelist} utilizes an encoder-decoder neural network to generate molecular structures de novo from tandem mass spectrometry, but its accuracy is less than 50\%. Comprehensive exploration of the diverse information embedded in these spectral modalities is crucial for advancing research in this domain.

\section{Chemistry Tools}
\label{sec:challenge3}

Although domian knowledge training and multi-modal enhancement can encode a certain amount of domain-specific knowledge into LLMs, it is constrained by scalability and intrinsic memory capacity \cite{chiang2024llamp}. In this section, We emphasize improving the capability of LLMs to tackle complex chemistry and embodied problems through the use of chemistry tools, such as operating experimental equipment for scientific research. 
We categorize these chemistry tools into three types: structured knowledge retrieval, machine learning (ML) models, and embodied robots.




\subsection{Structured Knowledge Retrieval}
\label{sec:structure_knowledge}


Structured knowledge retrieval, or retrieval-augmented generation (RAG, \cite{lewis2020retrieval}), has been proposed to alleviate hallucinations in both chemistry-specific and general LLMs \cite{xu2024hallucination}. The key component of knowledge retrieval is the knowledge source, and the retrieval method is typically determined by the source. 
We categorize common knowledge sources as follows:


\begin{enumerate}

    \item \textit{\textbf{Database}}: 
    There are many famous chemistry database, such as, Materials Project (MP, \citet{jain2013commentary}),  OPTIMADE~\cite{evans2024developments}. These databases cannot be accessed through direct web searches; instead, data retrieval requires following specific API documentation.
    LLaMP \cite{chiang2024llamp} design hierarchical ReAct \cite{yao2022react} agents that can dynamically and recursively interact with MP to ground LLMs on high-fidelity materials informatics.

    \item \textit{\textbf{Scientific Literature}}: 
    Peer-reviewed research articles are the most accurate and authoritative data source, and there are many Scholarly engines can help us find the related papers. 
    \citet{zheng2023chatgpt} propose to use ChatGPT for text mining the synthesis conditions of metal-organic frameworks (MOFs) and develop a ChatGPT Chemistry Assistant (CCA) chatbot base on the systhesis dataset and bibliographic context (such as authors and DOI), to alleviate hallucinatory errors.

    \item \textit{\textbf{Knowledge Graph}}: 
    A knowledge graph is a structured representation that allows for complex queries and provides insights that traditional databases cannot easily offer \cite{ye2024construction}. \citet{liu2024drak} propose KG-driven Knowledge Injection (DRAK-K) by retrieving the top-k most relevant pieces of knowledge and transforming the related knowledge into structured background context for LLMs.

\end{enumerate}

\subsection{ML Models}
\label{sec:ml_models}

LLMs are prone to worse than existing ML baselines \cite{guo2023can} in reaction-related tasks, and this tasks are difficult to be solved by knowledge retriveal.  
On the other hand, LLMs can interact with various tools (APIs) to accomplish complex tasks \cite{qin2023toolllm} in ReAct \cite{yao2022react} style , and we can boost chemistry LLMs performance with  SOTA ML models.
ChemCrow \cite{m2024augmenting} design reacttion tool set consist of  NameRXN, ReactionPredict and ReactionPlanner provied by RXN4Chemistry API from IBM Research, and plan the syntheses of an insect repellent and three organocatalysts. 
ChatChemTS \cite{ishida2024large} develop a user frendly chatbot named ChatChemTS which utilize AI-based molecule generators such as ChemTSv2 \cite{ishida2023chemtsv2} for molecular design. 
ChatMOF \cite{kang2024chatmof} foucs on generating new metal organic frameworks (MOFs, \citet{kitagawa2014metal}) which are useful in many chemical applications due to large porosity, high surface area,and exceptional tunability \cite{deng2012large}, and they also predict the properties of generated MOFs.
They adopt MOFTransformer \cite{kang2023multi} for the universal prediction of MOF properties and genetic algorithm \cite{park2022computational} to generate new MOFs, and achieve high accuracy of 95.7\% for predicting, and 87.5\% for generating tasks with GPT-4.

ML models can also help discover new catalyst by just giving feedback, ChemReasoner \cite{sprueill2024chemreasoner} use atomistic graph neural networks (GNNs) trained from quantum chemistry simulations for structure-based scoring, the GNNs are used to yeild reward and drive LLM towards catalysts with specific properties. This novel idea suggest that ML models not only can be used as tools aid in specific task, but also can be used as feeback to guide and stimulate the LLMs to fulfill the tasks by themselfs.

\subsection{Embodied Robots}
\label{sec:physical_environment}

Chemistry experiments are often resoure- and labor-intensive, and automated experiments canattain higher throughput and precision \cite{tom2024self}.
However, the discovery of new material requires not only automation but autonomy—the ability of an experimental agent to interpret data and make decisions based on it \cite{szymanski2023autonomous}, where LLMs are excellent at planing and reasoning, showing promise of sought-after system that autonomously designs and executes scientific experiments \cite{boiko2023autonomous}.

Coscientist \cite{boiko2023autonomous} is a GPT-4 driven AI system which can autonomously designs, plans and performs complex experiments, it demonstrate the versatility and performance across six tasks.
CLAIRify \cite{yoshikawa2023large} also leverage robots and LLM to automate chemistry experiments, and they pay more attention to how to generate syntactically valid programs in a data-scarce domain-specific language that incorporates environmental constraints.
ORGANA \cite{darvish2024organa} further extend CLAIRify with visual perception of the environment and support complex  experiments between multiple robots.

\section{Benchmarks}
\label{benchmarks_and_future}



Benchmarks are essential for evaluating the performance of chemistry LLMs on chemistry-related tasks and can be broadly categorized into two categories: science benchmarks and molecule-specific benchmarks.
Chemistry is a subset of science, and existing science benchmarks evaluate LLMs' ability to solve scientific problems, including those related to chemistry.
Existing science benchmarks, such as SciQ \cite{welbl2017crowdsourcing}, SciCode \cite{tian2024scicode}, ScienceQA \cite{lu2022learn}, AGIEval \cite{zhong2023agieval}, SciEval \cite{sun2024scieval}, SciBench \cite{wang2023scibench}, and VisScience\cite{jiang2024visscience}, typically cover a wide range of scientific disciplines, including biology, earth science, physics, chemistry, and even social science.
Although these science benchmarks include chemistry-related tasks, they are not specifically designed for chemistry and fail to address many chemistry-specific problems.

In contrast, molecule-specific benchmarks are designed to assess knowledge in molecule-related sciences (e.g., chemistry, materials science, biochemistry). 
ChemLLMBench \cite{guo2023can} first adapts traditional chemistry tasks to a language model setting, evaluating the performance of contemporary LLMs in zero-shot and few-shot prompts.
SciKnowEval \cite{feng2024sciknoweval} expands the chemistry domain to molecules by introducing a large dataset of 50,000 problems that assess various LLM abilities, including knowledge coverage, reflection and reasoning, and application.
MassSpecGym \cite{bushuiev2024massspecgym} focuses on characterization techniques, such as Tandem Mass Spectrometry (MS/MS), and evaluates the ability of LLMs to elucidate molecular structures from MS/MS data.
Notably, there are several other important chemistry benchmarks, including ScholarChemQA \cite{chen2024scholarchemqa}, SCIASSESS \cite{cai2024sciassess}, SciKnowEval \cite{feng2024sciknoweval}, ChemEVal \cite{huang2024chemeval}, \citet{alberts2024unraveling}, and MolPuzzles \cite{guo2024can}. Due to page limitations, we provide a brief overview of these benchmarks in Table \ref{tab:benchmarks}.

\section{Future Directions}
Although current approaches have made steady progress towards chemistry LLMs, there remains significant room for improvement. Future research directions can be categorized into three main aspects: data, model, and application.

\subsection{Data}

\paragraph{Data Diversity} 
Training data is the foundation of LLMs. However, most existing datasets are built from pre-existing sources, such as MoleculeNet \cite{wu2018moleculenet}, and cover a limited range of chemistry tasks. Future work should aim to create more diverse and comprehensive datasets to enhance the training of chemistry LLMs and broaden their capabilities.

\paragraph{CoT Reasoning} Chain-of-Thought (CoT, ~\citet{wei2022chain} ) reasoning is one of the most notable emergent abilities of LLMs, involving the generation of a sequence of intermediate steps leading to the final answer. However, existing chemistry LLMs often lack this critical reasoning capability due to simple training instruction pairs. Developing training data with explicit reasoning paths to effectively elicit the CoT ability in chemistry LLMs is a crucial direction for future research.

\paragraph{Chemical Modality} As described in Section \ref{sec:other_modal}, many chemistry-specific spectra are not yet fully exploited in in chemistry LLMs. 
However, these spectra contain rich structural information that can be valuable for various chemical tasks.
For example, tandem mass spectrometry (MS/MS) can provide detailed insights into the molecular structure, allowing for the identification and characterization of compounds and elucidation of reaction mechanisms.

\subsection{Model}
\paragraph{Multi-Modal Alignment}
Most works towards multi-modal chemistry LLMs always invole a single pair of modalities, limiting their representations ability. Align multiple N ( $\geq$ 3) modalities is a promising direction as different modalites are complementary and can provide more comprehensive understanding of chemistry molecules.

\paragraph{RLXF} RLHF is a crucial step in training powerful LLMs. Although obtaining human feedback is challenging, especially in chemistry where data annotation requires specialized domain knowledg, we can leverage advanced LLMs as assistants to guide this process. Additionally, we can also utilize results from professional chemistry software as a form of reward to align chemistry LLMs.

\subsection{Application}

\paragraph{Research Assistants}

Chemistry LLMs have the potential to serve as powerful research assistants, aiding chemists by automating routine tasks such as literature review, data analysis, and hypothesis generation. For future development, these models can be designed to understand complex scientific queries, provide insights from vast amounts of chemical literature, suggest experimental protocols, and even propose novel research directions. 

\paragraph{Automated Experimentation}

Automated experimentation is another promising direction for advancing chemistry LLMs. Integrating these models with automated laboratory systems can enable them to not only predict molecular properties or suggest potential chemical reactions but also design, execute, and analyze experiments in real-time. Future research should explore how chemistry LLMs can be trained and aligned to interact with automated experimental setups, ensuring reliability, safety, and compliance with scientific standards.



\section{Conclusion}
In this survey, we systematically investigate the current approaches to adapting general LLMs for chemistry LLMs. 
We highlight key challenges, including domain knowledge, multi-modal data, and the integration of chemistry-specific tools, and review existing efforts to address these challenges.
While significant progress has been made, achieving chemical general intelligence remains a distant goal, and we identify promising future directions. 
We hope this survey will inspire further innovative research in the field.

\section*{Limitations}
In this paper, a comprehensive review of existing methods for constructing chemistry-focused LLMs is presented, with an emphasis on three key aspects for enhancing general LLMs: domain-specific knowledge, multi-modal data, and chemistry tools. This survey aims to provide researchers with a concise understanding of chemistry LLMs and suggest potential directions for future research. However, certain limitations may be present.
\\
\\
\noindent \textbf{References.} Due to page limitations and the rapid development of the field, we may not include all relevant references and detailed technical information. However, we strive to keep our work up-to-date on our GitHub repository.

\section*{Acknowledgements}
I would like to express my gratitude to the anonymous reviewers for their meticulous and diligent review efforts.
This work was supported by National Science and Technology Major Project (Grant No. 2023ZD0120703), National Natural Science Foundation of China (Grant Nos. 92370206, U23B2057, 62120106006), and Shanghai Municipal Science and Technology Major Project (Grant No. 2021SHZDZX0102).

\bibliography{ref}

\appendix

\section{Related Work}\label{related_work}

The intersection of LLMs and chemistry is an urgent and rapidly growing field. Numerous works and reviews have addressed this topic, which can be broadly categorized into:

\subsection{General Science}

Several surveys focus on general science, including chemistry.
\citet{zhang2024comprehensive} explore LLM applications across mathematics, physics, biology, medicine, geography, geology, environmental science, and chemistry. However, the broad scope limits the depth of discussion on chemistry-specific LLMs. \citet{zhang2024scientific} focus more on the chemical domain but still include biological LLMs and BERT-style models, without discussing the emergent applications of chemistry-specific agents.

\subsection{Chemistry-Specific Surveys}

Chemistry's significance has drawn considerable attention, leading to various efforts summarizing current trends. 
\citet{xia2022systematic} review Chemical Pre-trained Models (CPMs) based on GNNs or Transformers but focus little on LLMs.
\citet{janakarajan2024language} emphasize the role of language models in molecular discovery but offer limited insights on training chemistry-specific LLMs. \citet{liao2024words} concentrate on molecule encoding and pretraining objectives, while \citet{pei2024leveraging} discuss progress from a multi-modal perspective, neglecting LLMs' tool-using potential. 
\citet{ramos2024review} review chemistry LLM agent applications in literature analysis, experiment planning, and hypothesis generation, but overlook multi-modal capabilities. 
Notably, these surveys categorize BERT-style LMs as LLMs, despite their need for task-specific fine-tuning and lack of emergent abilities.

\section{SFT Tasks Description}
\label{appendix:sft-task}
\begin{figure*}[]
\centering
\includegraphics[width=0.98\linewidth]{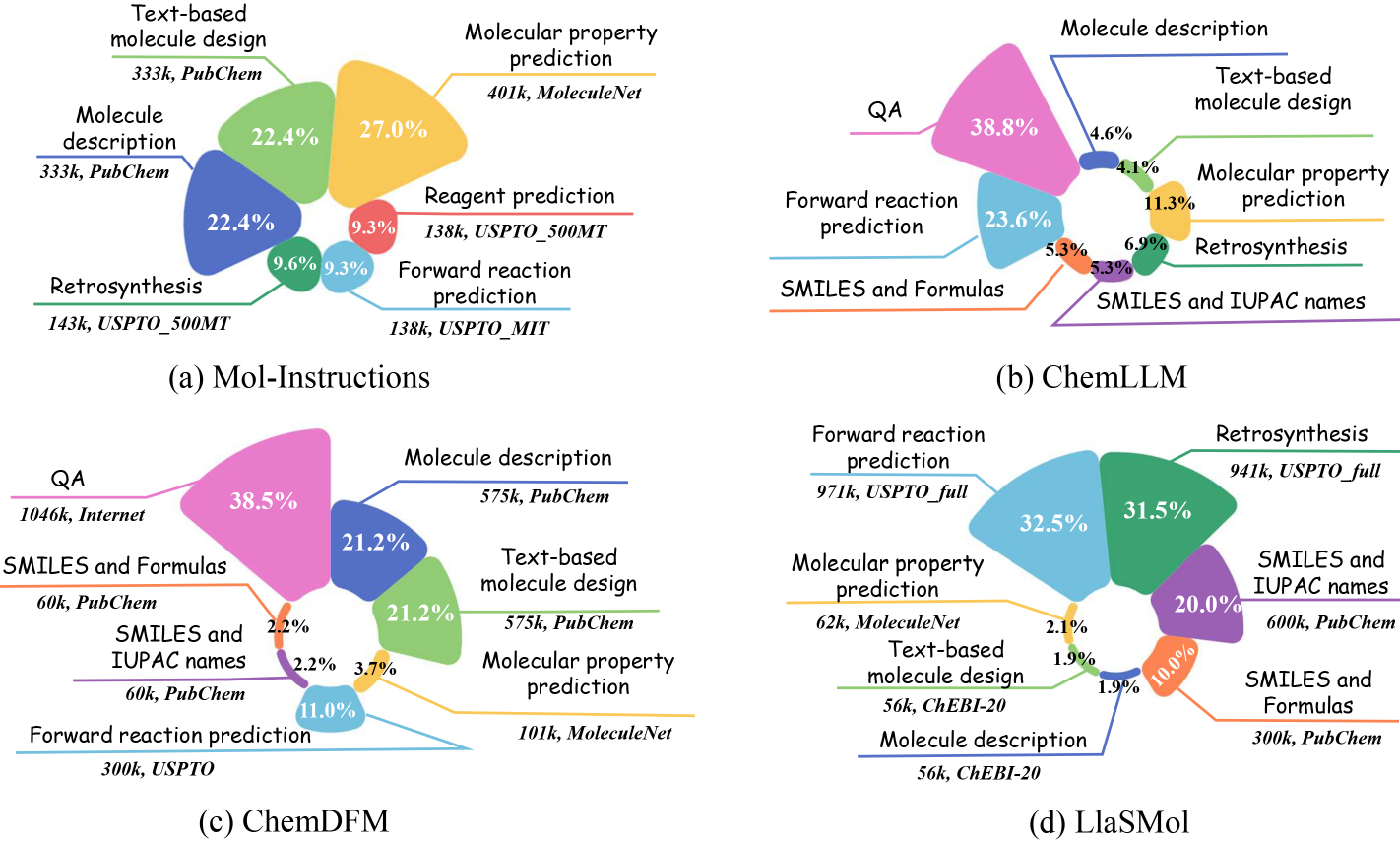} 
\caption{The compositional structure of representative SFT dataset. The definition of tasks above the the horizontal lines is shown in Table \ref{tab:sft-task}, the source and size of the different tasks are indicated below the horizontal lines, and percentages on the pie charts are present to show the difference of different dataset.}
\label{fig:sft-tasks}
\end{figure*}

The most frequently used chemistry tasks for SFT and their description are shown in Table \ref{tab:sft-task}. In accordance with the task division presented in Table \ref{tab:sft-task}, we illustrate in Figure \ref{fig:sft-tasks} the data distribution of the commonly used SFT dataset.

\begin{table*}[]
\centering
\scalebox{0.88}{
\begin{tabular}{@{}ccp{8cm}@{}}
\toprule
\textbf{Type}                                                                      & \textbf{Chemistry Tasks}      & \textbf{Description} \\ \midrule
\multirow{3}{*}{\begin{tabular}[c]{@{}c@{}}SMILES\\ Understanding\end{tabular}}    & Molecule description          &  Given a molecule SMILES, generating text description illuminating the structure, properties ,biological activity, and applications.                   \\ \cmidrule(l){2-3} 
                                                                                   & Text-based molecule design    &  Inverse task of molecule description, given a text description, generating the molecule SMILES.                    \\ \cmidrule(l){2-3} 
                                                                                   & Molecular property prediction &  Molecular property prediction focus on  drawn from Mquantum mechanics properties of molecules drawn from MoleculeNet.                    \\ \midrule
\multirow{3}{*}{\begin{tabular}[c]{@{}c@{}}Reaction \\ Understanding\end{tabular}} & Reagent prediction            &  Reagent prediction  generate suitable catalysts, solvents, or ancillary substances required for a specific chemical reaction.                   \\ \cmidrule(l){2-3} 
                                                                                   & Forward reaction prediction   &  Forward reaction prediction generate probable product(s) of a chemical reaction.                      \\ \cmidrule(l){2-3} 
                                                                                   & Retrosynthesis                &  Inverse task of forward reaction prediction, generate the synthesis routes and precursor molecules given target molecule.                \\ \midrule
\multirow{2}{*}{\begin{tabular}[c]{@{}c@{}}Notation \\ Alignment\end{tabular}}     & SMILES and IUPAC names        &  Given SMILES, generate IUPAC name, and reverse transformation.                 \\ \cmidrule(l){2-3} 
                                                                                   & SMILES and Formulas           &  Given SMILES, generate formulas, and reverse transformation.                     \\ \midrule
\begin{tabular}[c]{@{}c@{}}Chemistry-Related\\ QA\end{tabular}                     & QA                            &  Chemical QA extracted from existing dataset or exam.                    \\ \bottomrule
\end{tabular}
}
\caption{The most frequently used chemistry tasks for SFT.}
\label{tab:sft-task}
\end{table*}

\section{Molecule Image Dataset}
\label{appendix:molecule_image_dataset}
We describe the existing molecule image dataset in Table \ref{tab:Molecular-Image-dataset}.


\begin{table*}[h]
    \centering
    \begin{tabular}{>{\centering\arraybackslash}p{1.5cm} >{\centering\arraybackslash}p{5cm} >{\centering\arraybackslash}p{1cm} >{\centering\arraybackslash}p{7cm}}
        \toprule
        \textbf{} & \textbf{Dataset} & \textbf{scale} & \textbf{Description}\\
        \cline{1-4}
		\multirow{4}*{Synthetic}
        & USPTO-680K \cite{chen2024molnextr} & 680K & Multiple molecular formulas in one image\\ 
		\cline{2-4}
            & USPTO-30K \cite{morin2023molgrapher} & 30K 
            & {10K without bbreviation groups;
                10K has superatomic groups;
                10K is larger than 70 atoms}  \\
        \cline{2-4}
            & MolGrapher-Synthetic-300K \cite{morin2023molgrapher} & 300K & Rdkit generation \\
        \cline{2-4}
            & img2Mol \cite{clevert2021img2mol} & 41K & Rdkit generation \\
        \cline{2-4}
            & MMChemOCR \cite{li2024seeing} & 1K & closed source \\
        \cline{2-4}
            & MMCR-bench \cite{li2024seeing} & 1K & closed source \\
        \cline{2-4}
            & MMChemBench \cite{li2024seeing} & 700 & closed source \\
            
		\cline{1-4}
        \multirow{4}*{Realistic}
            & MolNexTR test data \cite{chen2024molnextr} & 18K & 5088 handwritten molecular images\\ 
        \cline{2-4}
            & RxnScribe \cite{qian2023rxnscribe} & 1413 & 4 forms of reaction images\\
        \cline{2-4}
            & OpenChemIED \cite{fan2024openchemie} & 254 & Only eval data is open source\\
        \cline{2-4}
            & ReactionDataExtractor 2.0 \cite{wilary2023reactiondataextractor} & 517 & Only eval data is open source\\
        \bottomrule
    \end{tabular}
    \caption{Overview of molecular image datasets, categorized into synthetic and realistic groups with details on their scale and descriptions. Synthetic datasets are primarily RDKit-generated or derived from large collections, while realistic datasets include handwritten and reaction images. Some datasets are closed-source or only provide evaluation data.}
    \label{tab:Molecular-Image-dataset}
\end{table*}

\section{Benchmarks}

We briefly introduce the existing benchmarks in Table \ref{tab:benchmarks}, covering aspects such as subject, task type, dynamics, source, and modality.

\begin{table*}[ht]
\centering
\scalebox{0.67}{
\begin{tabular}{@{}>{\centering\arraybackslash}p{6cm} >{\centering\arraybackslash}p{4cm} >{\centering\arraybackslash}p{2cm} >{\centering\arraybackslash}p{2cm} >{\centering\arraybackslash}p{2cm} >{\centering\arraybackslash}p{4cm}@{}}
\toprule
\textbf{Dataset} & \textbf{Subject} & \textbf{Task Type} & \textbf{Samples} & \textbf{Modality} & \textbf{Source}  \\ \midrule
SciQ \cite{welbl2017crowdsourcing}  & Bio, Chem, Earth, Phy    & MCQ, DA      & 1000      & Text         &      CK-12, OpenStax           \\

SciCode \cite{tian2024scicode}    &  Math, Phy, Chem, \newline Bio, Mat     & DA    & 338                &  Text       &  Research Paper                 \\

ScienceQA \cite{lu2022learn}    &  Natural, Social and Language Science   & MCQ   &    4,241             &   Image, Text      & School Curricula              \\ 

AGIEval  \cite{zhong2023agieval}    & Bio, Chem, Phy, Math, Law, \textit{at el.}    & MCQ,DA   &  8,062                &  Text       &       Human Exam             \\ 

SciEval  \cite{sun2024scieval}    & Bio, Chem, Phy    & MCQ,DA    &  15901               &  Text       &  Socratic QA , MedQA, PubMedQA        \\ 

SciBench \cite{wang2023scibench}    & Chem, Math, Phy    & DA   &      789      & Image, Text         &  TextBook               \\ 

VisScience \cite{jiang2024visscience}    &  Math, Chem, Phy    & MCQ,DA   & 3000                &  Image,Text       &     K12 education           \\ 

ChemLLMBench  \cite{guo2023can}    &  Chem    &  DA  &   800              &   Text     &      PubChem, MoleculeNet, USPTO, ChEBI,Suzuki           \\ 

SciKnowEval \cite{feng2024sciknoweval}    & Bio, Chem     &  MCQ, DA   &  50,048             &    Text     &   Literatures, Existing QAs, Databases   \\ 

MassSpecGym  \cite{bushuiev2024massspecgym}    &  Chem   &  DA  &     17,556            &   Spectra(Text)   &  MoNA, MassBank, GNPS,In-House Database       \\ 

ScholarChemQA  \cite{chen2024scholarchemqa}    &  Chem    &  MCQ   &      500           &   Text      &    Paper           \\ 

SciAssess  \cite{cai2024sciassess}    & Mat, Bio, Drug     & MCQ,DA   &   14,721              &   Image, Text      &    Existing benchmarks, Papers             \\ 

ChemEVal   \cite{huang2024chemeval}    &  Chem   &   DA  &       840      &   Text     &    Open-Source Data            \\ 

MolPuzzles   \cite{guo2024can}    &  Chem    &   DA &  19891         &  Spectra(Image)       & Textbook              \\ 

\citet{alberts2024unraveling}    &  Chem    &  DA  &    79K     &   Spectra(Text)      &       USPTO        \\ 
 
 \bottomrule
\end{tabular}
}
\caption{A brief introduction to the existing benchmarks. "MCQ" refers to Multi-Choice Questions, while "DA" denotes Direct-Answer tasks. "Samples" refers to the number of test set examples. The "Spectra" modality is distinctive, as spectra can be represented either as images or text.}
\label{tab:benchmarks}
\end{table*}

\end{document}